\documentclass[%
reprint,
 showpacs,
 amsmath,amssymb,
]{revtex4-1}
\usepackage{romannum}
\usepackage{graphicx}
\usepackage{dcolumn}
\usepackage{bm}

\usepackage{CJK}

\usepackage{color}
\usepackage{amssymb}
\usepackage{amsfonts}

\definecolor{Red}{rgb}{1,0,0}

\usepackage[colorlinks,urlcolor=blue,linkcolor=blue,citecolor=blue,anchorcolor=blue]{hyperref}
\makeatletter

\newcommand{\Rmnum}[1]{\expandafter\@slowromancap\romannumeral #1@}
\makeatother
\begin{document}

\preprint{APS/123-QED}

\title{Nonclassicality of fully degenerate triple-photon states and its application in generating non-Gaussian entanglement}


\author{Da Zhang$^{1,2}$}
\email{zhang1556433@sxnu.edu.cn}
\author{Yu Zhang$^1$}
\author{Juan Gao$^1$}

\affiliation{%
$^1$\mbox{School of Physics and Information Engineering, Shanxi Normal University, Taiyuan 03 31, China} \\
$^2$\mbox{Key Laboratory of Magnetic Molecules and Magnetic Information Materials, Ministry of Education, Taiyuan 03 31, China}
}%

\date{\today}

\begin{abstract}
\noindent
We theoretically demonstrate via numerical modeling that fully degenerate triple-photon states generated by three-mode spontaneous parametric down-conversion can be categorized into four distinct states: 0-phase, $\pi$/2-phase, $\pi$-phase, and 3$\pi$/2-phase squeezed states.
Using quantum relative entropy and Wigner negativity as quantitative measures, we show that the nonGaussianity and nonclassicality of these squeezed states increase with the increase of interaction strength.
Here, we investigate the correlation properties of the two-mode states produced by the interference of 0- and $\pi$-phase squeezed states at a beam splitter.
Employing the positive partial transpose criterion based on a high-order covariance matrix, we reveal the inseparability of these generated states at the 3rd- and 6th-order moments.
Notably, tuning the powers of the input light fields enables dynamic regulation of both this inseparability and quantum steering in the output modes.
The resulting two-mode state facilitates the remote preparation of two configurations of Wigner-negative states with tunable Wigner negativity.
Our results highlight the nonclassical nature of fully degenerate triple-photon states and establish a pathway for preparing non-Gaussian entanglement based on such states.
\end{abstract}

\maketitle
\section{Introduction}
Non-Gaussian states, defined by non-Gaussian Wigner function distributions, particularly those with negative regions, play an irreplaceable role in continuous-variable quantum information technology, as they circumvent the inherent limitations of the Gaussian framework \cite{vanloock.rmp.77.513.2005,weedrook.rmp.84.621.2012}.
Specifically, within the Gaussian paradigm, it is impossible to distill Gaussian entanglement via Gaussian operations alone \cite{eieret.prl.89.137903.2002}, and quantum information protocols relying solely on Gaussian states can be efficiently simulated by classical systems \cite{bartlett.prl.88.097904.2002,eisert.prl.109.230503.2012}.
Consequently, quantum states with Wigner negativity serve as indispensable resources for achieving quantum advantages in continuous-variable systems.
Moreover, using non-Gaussian entangled states as quantum channels can significantly enhance the fidelity of quantum teleportation \cite{olivares.pra.67.032314.2003,dell.pra.76.022301.2007}, the security threshold and information transmission rate in quantum key distribution protocols \cite{kumar.pra.100.052329,PhysRevA.102.012608}, and the measurement precision in quantum metrology \cite{PhysRevA.103.013705}.
Note that not all non-Gaussian states outperform their Gaussian counterparts in quantum information tasks, their utility depends on the specific protocol and nonclassical features required.

To date, one of the standard frameworks for preparing non-Gaussian states involves combining two-mode spontaneous parametric down-conversion (SPDC) with photon addition or subtraction \cite{agarwal1991nonclassical,ban1994quasicontinuous}.
It is shown that photon subtraction can not only enhance the entanglement degree of an input Gaussian state \cite{ourjoumtsev.prl.98.030502.2007} but also induce entanglement in initially separable states \cite{ourjoumtsev.np.5.189.2009}.
Building on this method, Menicucci et al. proposed a theoretical framework for continuous-variable universal quantum computation \cite{nielsen.prl.97.110501.2006}, while Serikawa et al. experimentally demonstrated entanglement distillation with Gaussian inputs via local photon subtraction \cite{masahide.np.4.178.2010}.
Besides, laboratory experiments have realized cat states \cite{ourjoumtsev.science.312.83.2006,sychev.np.11.379.2017,furusawa.prl.121.143602.2018,nicola.prl.124.033604.2020}, superpositions of Fock states \cite{lvovsky.np.4.2010,furusawa.oe.5.2013}, hybrid entangled states \cite{jeong.np.8.564.2014,morin.np.8.570.2014}, and multimode non-Gaussian states \cite{ra.np.non.2020}.

As another deterministic framework for preparing non-Gaussian states, three-mode SPDC has recently been realized in superconducting circuits \cite{chang.prx.10.011011.2020}.
By tuning the interaction time to enhance the interaction strength, optical fibers \cite{cavanna.pra.101.033840.2020} and waveguides \cite{Moebius.oe.9.9932.2016} are emerging as promising physical platforms for implementing this process.
Depending on the degeneracy of the three modes, the states generated by three-mode SPDC can be classified into three categories: fully degenerate, partially degenerate, and nondegenerate triple-photon states (TPS).
There is a quantum correlation between the linear quadratures of the nondegenerate mode and the 2nd-order quadratures of the degenerate mode in the partially degenerate TPS, so it is also referred to as nonlinear entangled states \cite{lam.prl.114.100403.2015,zhang.ctp.75.205101.2023}.
Their entanglement properties can be fully characterized by a series of $3n$-order covariance matrices (CMs) \cite{zhang.prl.127.150502.2021}, which capture higher-order statistical dependencies.
In contrast, nondegenerate TPS are inherently non-Gaussian Greenberger-Horne-Zeilinger states with super-Gaussian statistics \cite{zhang.pra.013704.2021}, demonstrating full inseparability and genuine tripartite entanglement in $3n$-order moments \cite{zhang.PRL.130.093602.2023,agust.prl.125.020502,tian.prapplied.18.024065.2022}.
Although the quantum properties of fully degenerate TPS, which exhibits Wigner negativity, have been studied in some Refs. \cite{felbinger.prl.80.492.1998,kamel.crp.8.206.2007}, the veil covering it has not yet been completely lifted.
\begin{figure*}[htpb]
\centering
  \includegraphics[width=14.5cm]{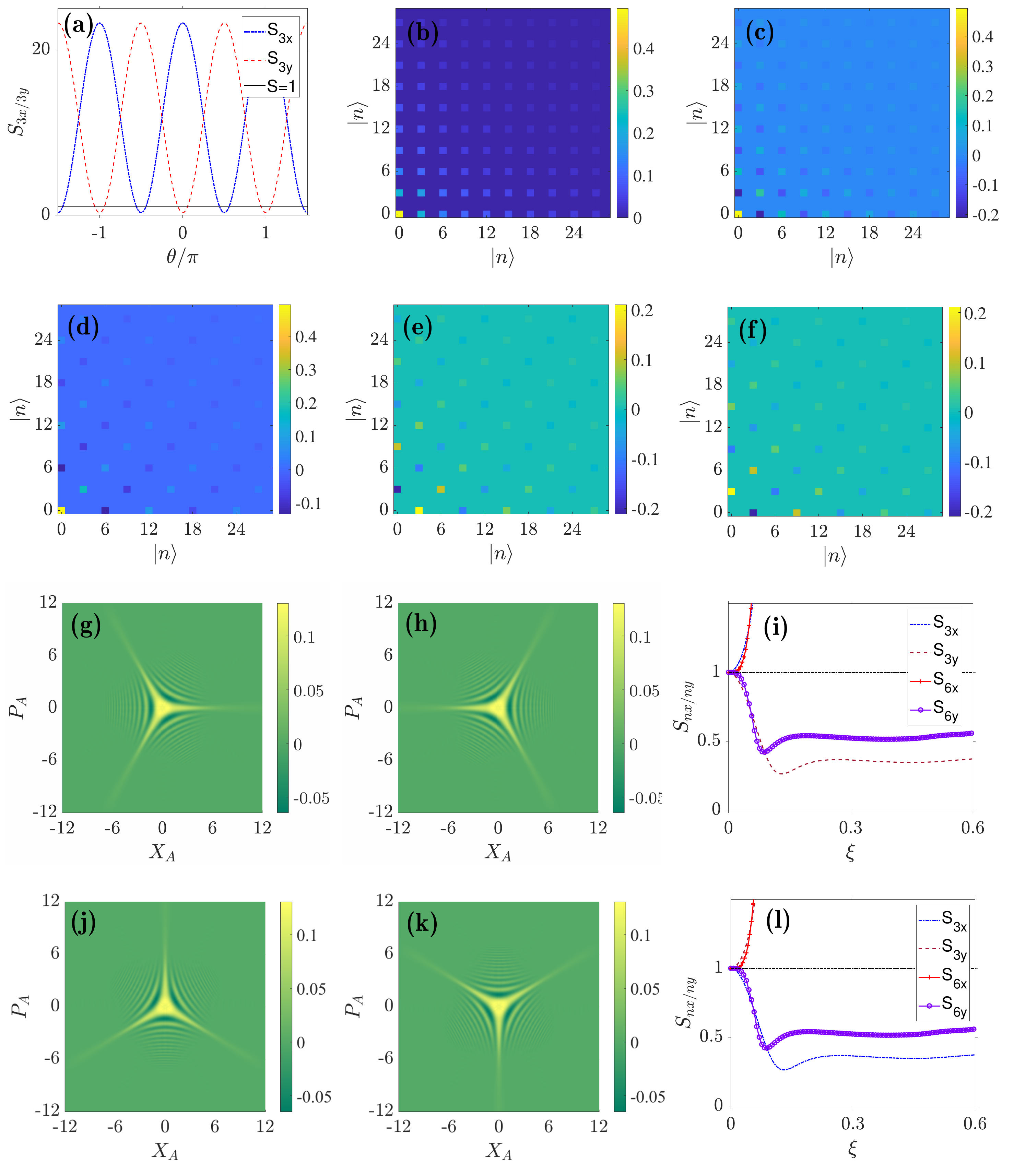}  
  \caption{(a) Evolution of $S_{3x}$ and $S_{3y}$ with interaction strength $\xi=\kappa\alpha_pt=0.5$, where $\alpha_p=5$. (b), (c), and (d) show the real parts of the density matrices for the 0-phase, $\pi$-phase, and $\pi/2$-phase squeezed states, respectively. (e) and (f) depict the imaginary parts of the density matrices for the $\pi/2$-phase and $3\pi/2$-phase squeezed states, respectively. (g), (h), (j), and (k) present the Wigner functions of the 0-phase, $\pi$-phase, $\pi/2$-phase, and $3\pi/2$-phase squeezed states, respectively. (i) and (l) show the evolution of $S_{ko}$ versus $\xi$ for the 0- and $\pi/2$-phase squeezed state at 3rd- and 6th-order moments, respectively.}
  \label{fig1}
\end{figure*}
Inspired by the generation of two-mode Gaussian states via beam splitter interference of amplitude- and phase-squeezed states, we are curious whether two-mode non-Gaussian entangled states can be prepared using fully degenerate TPS and beam-splitter operations.

In this work, we show that the 3rd-order amplitude and phase quadrature operators of fully degenerate TPS exhibit alternating squeezing and antisqueezing as the pump phase is tuned.
By tuning the pump phase to different values, we characterize four distinct fully degenerate TPS, hereafter referred to as 0-phase, $\pi$/2-phase, $\pi$-phase, and 3$\pi$/2-phase squeezed states in sequence.
We quantify the nonGaussianity and nonclassicality of the system using quantum relative entropy and Wigner negativity, respectively, and find that they both increase with increasing interaction strength.
Bisides, we study the generation of two-mode non-Gaussian states by interfering 0-phase and $\pi$-phase squeezed states at a beam splitter.
By applying the positive partial transpose (PPT) criterion to the higher-order CM, we reveal the entanglement properties of the resulting state in terms of 3rd- and 6th-order moments.
Furthermore, based on the Hillery-Zubairy steering criterion, we characterize the quantum steering properties of the two output modes at the 3rd- and 6th-order moments.
By tuning the intensities of the input 0- and $\pi$-phase squeezed light fields, we find that both the entanglement and quantum steering of the two output modes at the 3rd- and 6th-order moments can be dynamically regulated.
Finally, we utilize the generated two-mode non-Gaussian states to enable the remote preparation of Wigner-negative states with tunable negativity.
Our results establish a pathway for the deterministic preparation of non-Gaussian entangled states using these non-Gaussian squeezed states as building blocks.

The paper is structured as follows. In Section \Romannum{2}, we classify fully degenerate TPS and analyze their nonclassical properties such as higher-order squeezing and Wigner negativity.
Section \Romannum{3} investigates a typical scenario for generating two-mode non-Gaussian entangled states using fully degenerate TPS.
Using the PPT and Hillery-Zubairy steering criteria respectively, we reveal their inseparability and steerability at 3rd- and 6th-order moments.
Besides, we explore their application in remotely preparing Wigner-negative states.
The conclusions are presented in Section \Romannum{4}.
\section{Nonclassicality of fully degenerate TPS}
We initiate our analysis with the interaction Hamiltonian \cite{chang.prx.10.011011.2020}
\begin{align}\label{eq1}
\hat{H}&=i\hbar\kappa(\hat{a}^{\dag 3}\hat{p}e^{-i\theta}-\hat{a}^3\hat{p}^{\dag}e^{i\theta}),
\end{align}
describing the fully degenerate three-mode SPDC, where $\kappa$ is the third-order coupling constant describing the strength of the nonlinear interaction and $\theta$ represents the pump phase.
The annihilation operators $\hat{a}$ and $\hat{p}$ denote the down-conversion mode and the pump mode, respectively.
Using this Hamiltonian, the final state of the system at time $t$ can be derived by solving the Schr$\mathrm{\ddot{o}}$dinger equation, under the assumption that the initial state is vacuum for the triplets and a coherent mode $\alpha_p$ for the pump.

We define the $k$th-order amplitude and phase quadrature operators $\hat{x}^{k}=(\hat{a}^{k}+\hat{a}^{\dag k})/2$ and $\hat{y}^{k}=i(\hat{a}^{\dag k}-\hat{a}^{k})/2$, satisfying the commutation relation $[\hat{x}^{k},\hat{y}^{k}]=i\hat{f}^k$, where $k$ is a positive integer and the expression of the operator $\hat{f}^k$ is provided in the Appendix.
By introducing the rotation operator $\hat{U}_0(\theta)=\exp(i\hat{H}_0\theta/\hbar)$ defined based on the intrinsic Hamiltonian $\hat{H}_0$ of the system, it is shown theoretically in \cite{hillery.pra.36.3796.1986} that the fully degenerate TPS generated in the spontaneous parameter regime has only nonzero moments of order $3k$, that is, $\langle\hat{x}^{n}\rangle=\langle\hat{y}^{n}\rangle=0$ when $n\neq3k$.
If the state is squeezed in the moments of order $k$, then $S_{ko}=2(\langle(\hat{o}^{k})^2\rangle-\langle\hat{o}^{k}\rangle^2)/\langle\hat{f}^k\rangle<1$ ($o=x,y$).
Figure \ref{fig1}(a) illustrates the evolution of $S_{3x/3y}$ as a function of the pump phase $\theta$ for $\kappa\alpha_pt=0.5$, where $\alpha_p=5$.
The squeezing of the 3rd-order phase quadrature operator is maximized at $\theta=0$ and $\theta=\pm\pi$, whereas the squeezing of the 3rd-order amplitude quadrature operator is maximized at
$\theta=\pi/2$ and $\theta=3\pi/2$.

Considering the three-fold symmetry of fully degenerate TPS in Wigner space \cite{kamel.crp.8.206.2007}, these squeezed states generated at different pump phases exhibit distinct properties, which differ from the amplitude and phase squeezed states produced by two-mode SPDC.
Specifically, the three-fold symmetry implies that the Wigner function remains invariant under a 120$^o$ rotation.
However, when the squeezing of the 3rd-order phase quadrature operator reaches its maximum, the difference between adjacent $\theta$ values is $\pi$, which is equivalent to a 60$^o$ rotation of mode $\hat{a}$ in Wigner space, apparently contradicting the three-fold symmetry.
The same analysis applies to cases of $\theta=\pi/2$ and $\theta=3\pi/2$.
This means that the fully degenerate TPS are different when the pump phase $\theta$ takes different values.
To distinguish these different squeezed states, we classify them as 0-phase, $\pi$/2-phase, $\pi$-phase, and 3$\pi$/2-phase squeezed states according to the value of the pump phase.

Figure \ref{fig1}(b) and \ref{fig1}(c) respectively display the density matrices of the 0-phase and $\pi$-phase squeezed states.
As expected, the photon-number distribution of the $0$-phase squeezed state only exists in $3k$ and its off-diagonal elements are all positive, indicating a zero phase difference across different $3k$ values.
In contrast, the $\pi$-phase squeezed state features a relative phase difference of $\pi$ between adjacent photon number.
The real parts of the density matrices for the $\pi/2$-phase and $3\pi/2$-phase squeezed states are identical and shown in Fig. \ref{fig1}(d), while their imaginary parts are presented in Fig. \ref{fig1}(e) and \ref{fig1}(f), respectively.
The adjacent photon number in the $\pi/2$-phase ($3\pi/2$-phase) squeezed state exhibit relative phase differences of $\pi/2$ ($3\pi/2$).
Figures \ref{fig1}(g), \ref{fig1}(h), \ref{fig1}(j), and \ref{fig1}(k) depict the Wigner functions of the 0-phase, $\pi$-phase, $\pi/2$-phase, and $3\pi/2$-phase squeezed states, respectively.
All exhibit star-shaped profiles with threefold rotational symmetry and pronounced negative regions.
In particular, the 0-phase squeezed state is rotated by 30$^o$ counterclockwise (clockwise) in the Wigner space to generate the $3\pi/2$ ($\pi/2$)-phase squeezed states.
This is also easy to understand.
Rotating mode $\hat{a}$ by 30$^o$ in Wigner space induces a 90$^o$ phase accumulation in the $\hat{H}$ due to its three-fold degeneracy.

Figure \ref{fig1}(i) shows the evolution of $S_{ko}$ for the 0-phase squeezed state as a function of $\xi$.
Initially, phase squeezing emerges at the 3rd-order moments.
As $\xi$ increases, phase squeezing at the 6th-order moments becomes prominent, yet it remains weaker than that at the third order.
Figure \ref{fig1}(l) is the same as Fig. \ref{fig1}(i) but for $\pi/2$-phase squeezed state.
Unlike the 0-phase squeezed state where squeezing is confined to $3n$th-order phase quadratures, the $\pi/2$-phase squeezed state exhibits amplitude squeezing at the 3rd-order moments and phase squeezing at the 6th-order moments.
The squeezing behavior of the $\pi$-phase ($\pi/2$-phase) squeezed state is identical to that of the 0-phase ($3\pi/2$-phase) squeezed state.

\begin{figure}[htpb]
\centering
  \includegraphics[width=6cm]{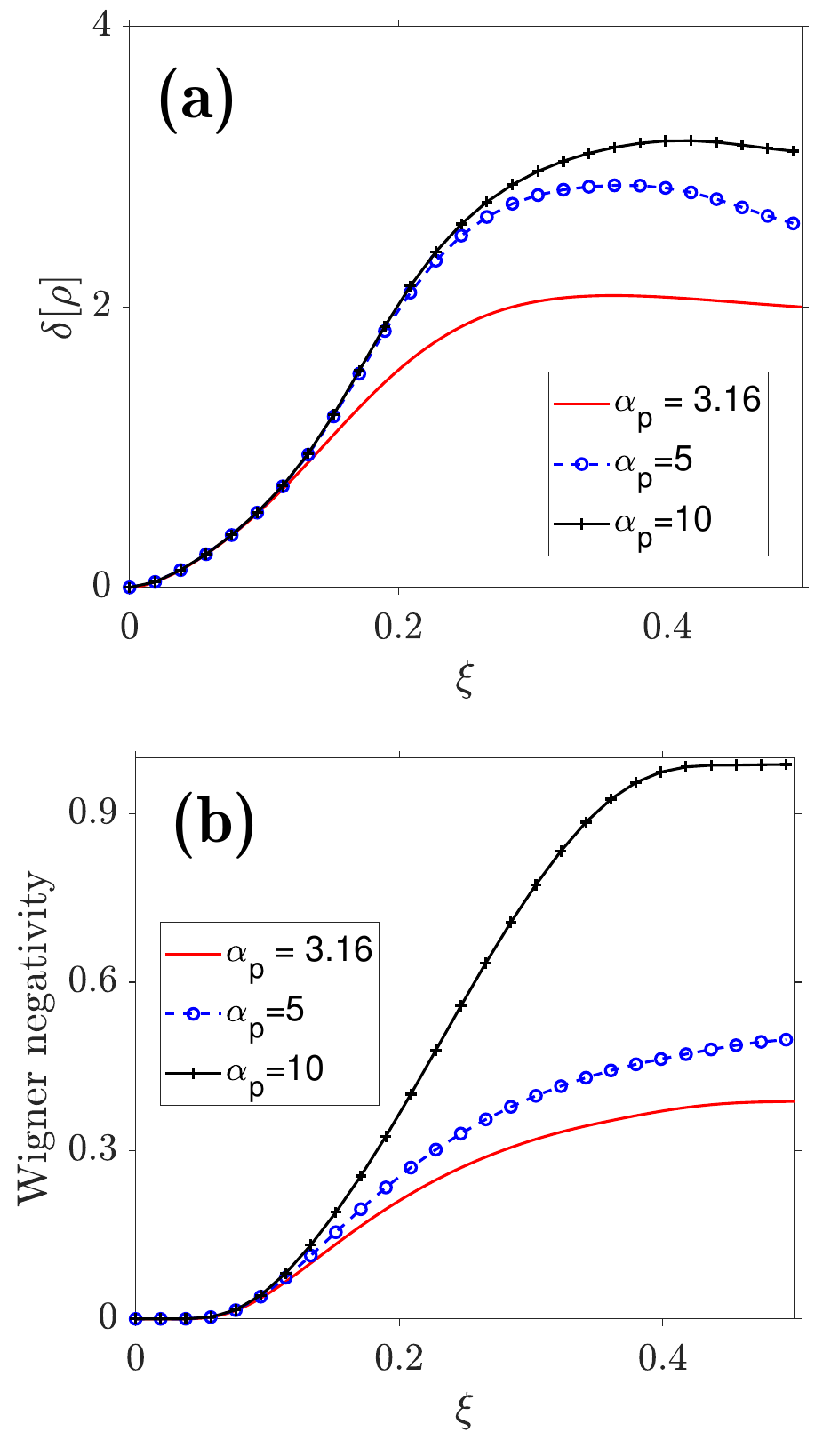}  
  \caption{Evolution of quantum relative entropy $\delta[\rho]$ and Wigner negativity $\mathcal{W}$ versus $\xi$ under different pump amplitudes.}
  \label{fig2}
\end{figure}
As a class of non-Gaussian states, we now use quantum relative entropy to quantify the evolution of their non-Gaussianity with respect to the interaction strength.
The quantum relative entropy between a given quantum state $\rho$ and a reference quantum state $\gamma$ is defined as \cite{marco.pra.78.6.2008}
\begin{align}\label{eq2}
\delta[\rho]=E_{vN}(\gamma)-E_{vN}(\rho),
\end{align}
where $E_{vN}(\rho)=\textrm{Tr}[\rho\ln\rho]$ denotes the von Neumann entropy, and the states $\gamma$ and $\rho$ share identical covariance matrix and first-order moments.
The von Neumann entropy of a Gaussian state with covariance matrix $\gamma$ and symplectic eigenvalue $\nu_j$ is
\begin{align}\label{eq3}
E_{vN}(\gamma)=\sum_{j=1}^n f(\nu_j),
\end{align}
where $f(x)=(x+\frac{1}{2})\ln(x+\frac{1}{2}) - (x-\frac{1}{2})\ln(x-\frac{1}{2})$.
Figure \ref{fig2}(a) shows the evolution of $\delta[\rho]$ for the fully degenerate TPS as a function of $\xi$ under different pump amplitudes.
We can see that $\delta[\rho]$ increases with the increase of $\xi$.
Notably, a higher pump amplitude leads to a greater value of $\delta[\varrho]$.
This phenomenon arises because a brighter pump mitigates pump depletion, driving the process closer to the parametric approximation regime \cite{PhysRevA.47.1237}.

We next characterize the nonclassicality of these phase-squeezed states using Wigner negativity $\mathcal{W}$, defined as
\begin{align}\label{eq4}
\mathcal{W}_A&=\frac{1}{2}\int [|W(X_A,P_A)|-W(X_A,P_A)]dX_AdP_A,
\end{align}
where $W(X_A,P_A)$ is the Wigner function of mode $\hat{a}$, $X_A$ and $P_A$ are phase-space variables.
Figure \ref{fig2}(b) illustrates the evolution of $\mathcal{W}_A$ versus $\xi$.
With the increase of $\xi$, the state gradually evolves from the vacuum with circular symmetry to the fully degenerate TPS with three-fold symmetry,
and the negativity of $W(X_A,P_A)$ begins to appear when $\xi$ reaches a certain value.
After that, the brighter the pump, the larger the $\mathcal{W}_A$.
Remind that when $\xi$ is the same, these different squeezed states have the same $\delta[\rho]$ and $\mathcal{W}_A$ due to rotational symmetry.
\begin{figure*}[htpb]
\centering
  \includegraphics[width=16cm]{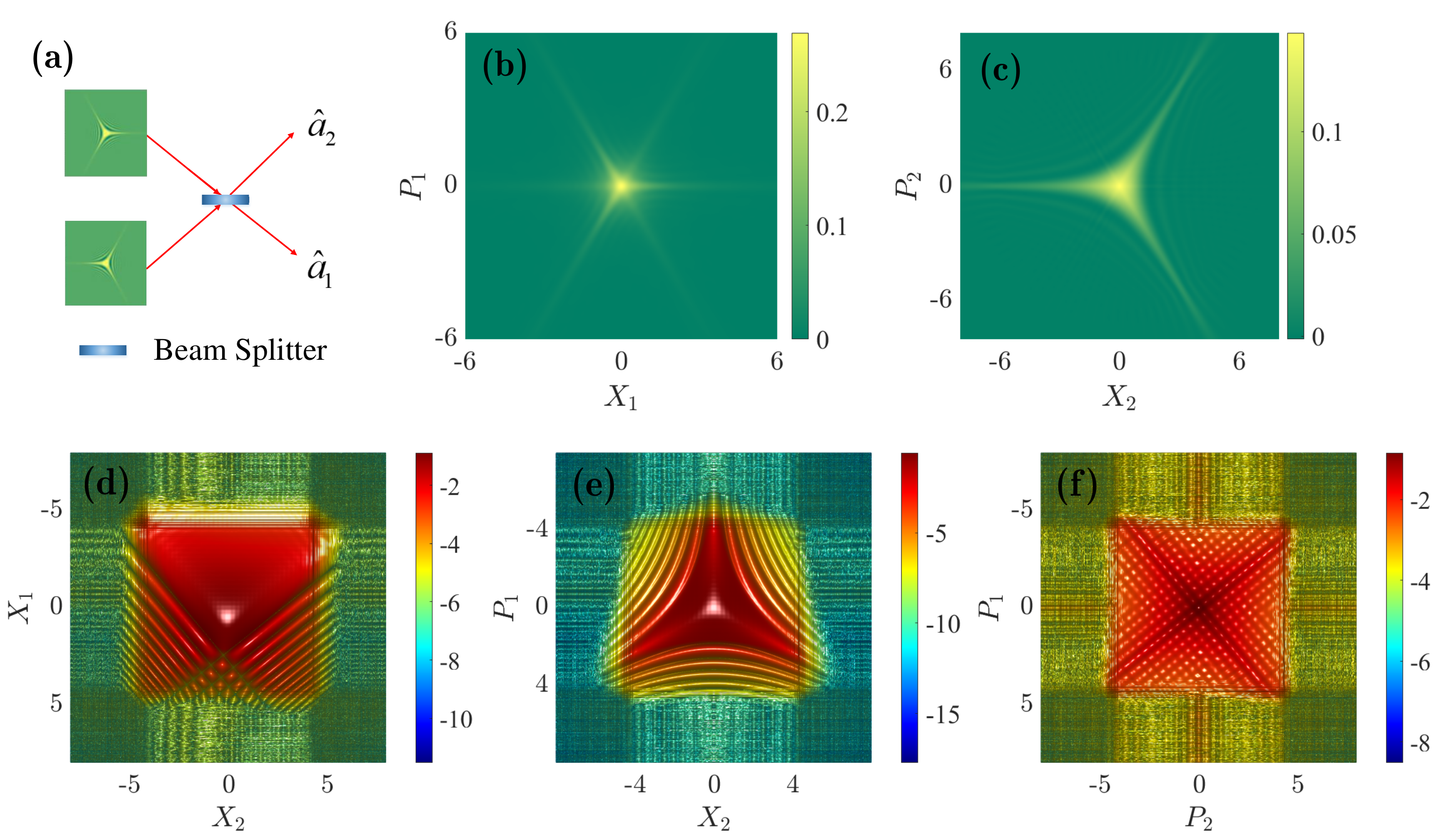}  
  \caption{(a) Schematic diagram of the preparation of a two-mode non-Gaussian state, where the 0-phase and $\pi$-phase squeezed states are mixed on a beam splitter with transmittance 0.6. Wigner functions for modes $\hat{a}_1$ (b) and $\hat{a}_2$ (c) when $\xi_0=\xi_{\pi}=0.2$. Standard quadrature joint probability distribution for (d) $X_1$ and $X_2$, (e) $P_1$ and $X_2$, and (f) $P_1$ and $P_2$ when $\xi_0=\xi_{\pi}=0.1$.}
  \label{fig3}
\end{figure*}
\section{Entanglement properties of two-mode non-Gaussian state}
In analogy to the amplitude- and phase-squeezed states in the Gaussian framework, we are curious whether these four non-Gaussian squeezed states can serve as fundamental building blocks to unconditionally prepare two-mode non-Gaussian states.
We consider a typical example as illustrated in Fig. \ref{fig3}(a), where a 0-phase and a $\pi$-phase squeezed state interfere on a beam splitter with transmittance 0.6, yielding two output modes $\hat{a}_1$ and $\hat{a}_2$.
Here we first assume that the interaction strengths for generating the two input squeezed states are the same, that is, $\kappa_ 0=\kappa_{\pi}$, where $\kappa_{0/\pi}$ is the nonlinear coupling constant for generating the 0$/\pi$-phase squeezed state.
Figure \ref{fig3}(b) shows the Wigner function of output mode $\hat{a}_1$ when $\xi=0.2$.
Unlike the input fully degenerate TPS, which features three-fold symmetry, mode $\hat{a}_1$ displays a snowflake-like profile that features six-fold symmetry.
In contrast, output mode $\hat{a}_2$ retains three-fold symmetry, as depicted in Fig. \ref{fig3}(c).
These two distinct distributions reveal an inherent asymmetry between the two output modes.
Figures \ref{fig3}(d), \ref{fig3}(e), and \ref{fig3}(f) respectively show the joint probability distributions $X_1-X_2$,  $P_1-X_2$, and $P_1-P_2$.
These intricate distributions imply complex quantum correlation properties between modes $\hat{a}_1$ and $\hat{a}_2$.

In the Heisenberg picture, using the high-order quadrature operators as the observables, the quantum correlation between the two output modes can be described as
\begin{align*}
&\langle\hat{x}^n_1\hat{x}^k_2\rangle=\sum^{n}_{n_1=0}\sum^{k}_{k_1=0}N_{n_1}K_{k_1}\mathrm{Tr}[\hat{a}^{\dag n-n_1+k_1}_0\hat{a}^{\dag k-k_1+n_1}_{\pi}\rho_0\rho_{\pi}] \nonumber \\
&+\sum^{n}_{n_1=0}\sum^{k}_{k_2=0}N_{n_1}K_{k_2}\mathrm{Tr}[\hat{a}^{\dag n-n_1}_0\hat{a}^{k_2}_0\hat{a}^{\dag n_1}_{\pi}\hat{a}^{k-k_2}_{\pi}\rho_0\rho_{\pi}]       \nonumber \\
&+\sum^{n}_{n_2=0}\sum^{k}_{k_1=0}N_{n_2}K_{k_1}\mathrm{Tr}[\hat{a}^{n-n_2}_0\hat{a}^{\dag k_1}_0\hat{a}^{n_2}_{\pi}\hat{a}^{\dag k-k_1}_{\pi}\rho_0\rho_{\pi}]       \nonumber \\
\end{align*}
\begin{align}\label{eq5}
&+\sum^{n}_{n_2=0}\sum^{k}_{k_2=0}N_{n_2}K_{k_2}\mathrm{Tr}[\hat{a}^{n-n_2+k_2}_0\hat{a}^{k-k_2+n_2}_{\pi}\rho_0\rho_{\pi}]
\end{align}
where $N_{n_i}=C_n^{n_i}0.6^{(n-n_i)/2}0.4^{n_i/2}$ ($K_{k_i}$) is a trivial constant.
Here, $\rho_0$ and $\rho_{\pi}$ represent the density matrices of the 0-phase and $\pi$-phase squeezed states, respectively.

Drawing upon the fact that the fully degenerate TPS only has non-zero moments of order $3k$, it follows that $\mathrm{Tr}[a^n_{0/\pi}\rho_{0/\pi}]=0$ when $n\neq3k$.
Starting from this premise, one can deduce that the conditions $n-n_1+k_1=3m_1$ and $k-k_1+n_1=3m_2$ must hold when the first term on the right-hand side of Eq. (\ref{eq5}) to be nonvanishing.
Similar conclusions can be drawn by analyzing the remaining three terms in Eq. (\ref{eq5}).
Straightforward mathematical analysis reveals that $\langle\hat{x}^n_1\hat{x}^k_2\rangle\neq0$ when $n\pm k=3m$ or $n=k$, thereby highlighting the intricate correlation features between modes $\hat{a}_1$ and $\hat{a}_2$.

Based on the foregoing analysis, we now use the PPT criterion applied to high-order CMs to diagnose the entanglement properties of the generated state.
Higher-order quadrature operators are first assembled into vectors $\hat{r}^{kl}=(\hat{x}^k_1,\hat{y}^k_1,\hat{x}^l_2,\hat{y}^l_2)^T$, which satisfy the commutation relations
\begin{align}\label{eq6}
[\hat{r}^{kl}_i,\hat{r}^{kl}_j]=i\Omega^{kl}_{ij},
\end{align}
where $\Omega^{kl} =\text{diag}(J_{k},J_{l})$, $J_{k}=\text{adiag}(\hat{f}^{k},-\hat{f}^{k})$, and $i,j=1,\dots,4$.
The higher-order CM is defined as $V^{kl}_{ij}=\langle\Delta\hat{r}^{kl}_i\Delta\hat{r}^{kl}_j+\Delta\hat{r}^{kl}_j\Delta\hat{r}^{kl}_i\rangle/2$, where
$\Delta\hat{r}^{kl}=\hat{r}^{kl}-\langle \hat{r}^{kl}\rangle$ and $\langle \hat{r}^{kl}\rangle=\mathrm{tr}[\hat{r}^{kl}\hat{\rho}]$, with $\hat{\rho}$ being the density operator of the system.
By inserting the commutation relations (\ref{eq6}), it is shown that the higher-order CM satisfies the uncertainty principle
\begin{align}\label{eq7}
V^{kl}+\frac{i}{2}\langle\Omega^{kl}\rangle \geq 0.
\end{align}

If $V^{kl}$ is separable, the partially transposed $\widetilde{V}^{kl}$ must still satisfy
\begin{align}\label{eq8}
\widetilde{V}^{kl}+\frac{i}{2}\langle\Omega^{kl}\rangle \geq 0.
\end{align}
Violation of inequality (\ref{eq8}) is equivalent to $\tilde{\nu}_{kl-}<0$, where $\tilde{\nu}_{kl-}$ is the smallest eigenvalue of $\widetilde{V}^{kl}+i\langle\Omega^{kl}\rangle/2$, suggesting the existence of entanglement.
It is important to note that a critical prerequisite for the PPT criterion established in \cite{zhang.prl.127.150502.2021} to serve as a necessary and sufficient condition for the separability of the two-mode nonlinear quantum state $V^{kl}$ is that the quantum correlations of this state at the $(k+l)$-th moments can be fully characterized by $V^{kl}$.
However, the two-mode non-Gaussian state generated in Fig. \ref{fig3}(a) fails to satisfy this condition, as evidenced by the preceding analysis of its correlation properties.

\begin{figure*}[htpb]
\centering
  \includegraphics[width=17.5cm]{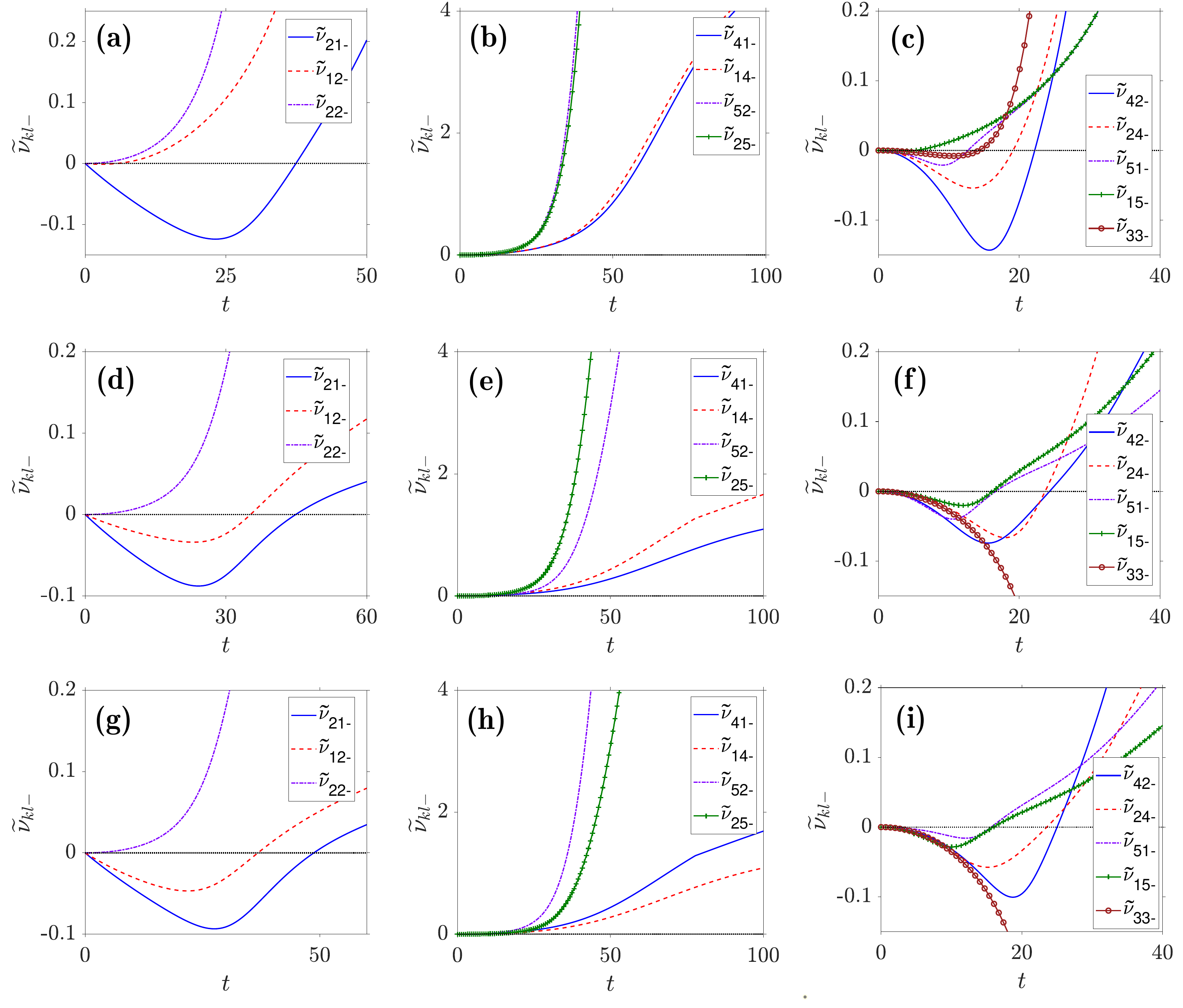}  
  \caption{ (a), (b), and (c) show the evolution of $\tilde{\nu}_{kl-}$ versus $t$ for different $k$ and $l$ when $\kappa_ 0=\kappa_{\pi}=0.0008$ and $\alpha_p=5$. $\tilde{\nu}_{kl-}<0$ implies the presence of entanglement at ($k+l$)-order moments. (d), (e) and (f) are analogous to (a), (b) and (c), but with $\kappa_ 0=4\kappa_{\pi}=0.0008$, while (g), (h) and (i) correspond to $4\kappa_ 0=\kappa_{\pi}=0.0008$.}
  \label{fig4}
\end{figure*}

Figure \ref{fig4}(a) shows the evolution of $\tilde{\nu}_{21-}$, $\tilde{\nu}_{12-}$, and $\tilde{\nu}_{22-}$ as a function of $t$.
In part of the parameter interval, $\tilde{\nu}_{21-}$ is less than zero, which means that the two modes are entangled in the 3rd-order moments.
This entanglement property, specifically, the entanglement between the 2nd-order quadrature of mode $\hat{a}_1$ and the linear quadrature of mode $\hat{a}_2$, is similar to the entanglement exhibited by the state generated by Hamiltonian $H_I=i\hbar\kappa\hat{a}_1^{\dag 2}\hat{a}_2^{\dag}\hat{p}+H.c.$ in the spontaneous parametric regime at the 3rd-order moments \cite{zhang.prl.127.150502.2021}.
It should be noted that $\tilde{\nu}_{12-}$ is greater than  0 in the entire parameter interval, which means that the modes $\hat{a}_1$ and $\hat{a}_2$ are asymmetric.
In addition, although the two output modes are also correlated in the 4th-order moments, $\tilde{\nu}_{22-}>0$, so we can not get any conclusion about the entanglement.
The same is true for the CMs of order 5 and 7, as shown in Fig. \ref{fig4}(b).
Figure \ref{fig4}(c) demonstrates the evolution of $\tilde{\nu}_{kl-}$ versus $t$ based on the 6th-order moments.
We can see that $\tilde{\nu}_{42-}$, $\tilde{\nu}_{24-}$, $\tilde{\nu}_{51-}$, and $\tilde{\nu}_{33-}$ are all less than zero in some parameter intervals, which implies that there are four different types of entanglement between the two output modes at the 6th-order moments.

\begin{figure*}[htpb]
\centering
  \includegraphics[width=17.5cm]{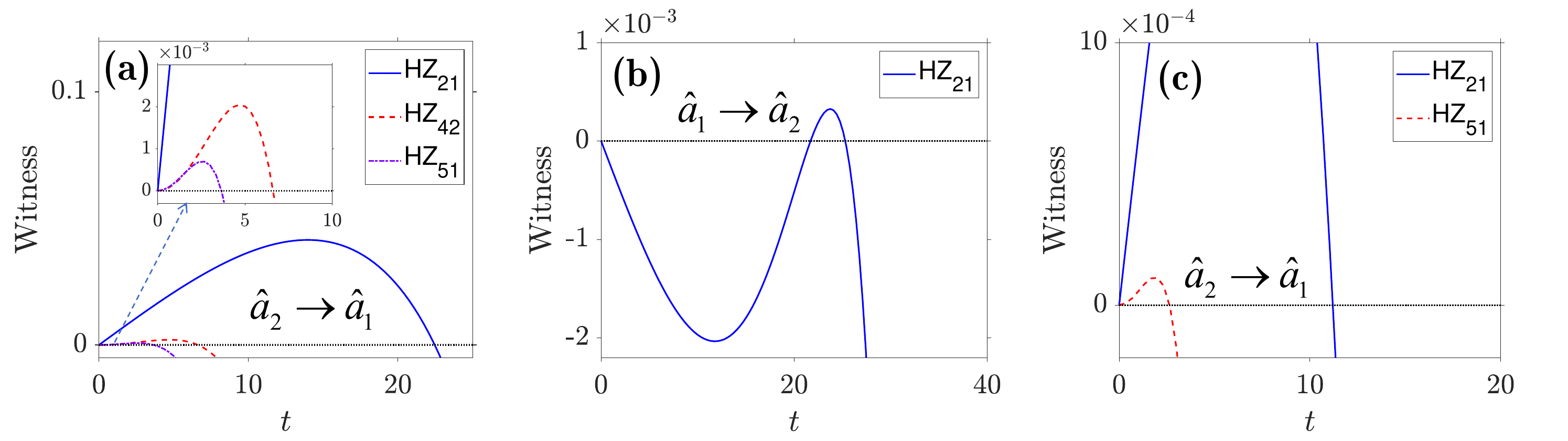}  
  \caption{(a) Evolution of the HZ$_{kl}$ with $t$ when $\kappa_0=\kappa_{\pi}=0.0008$. (b) and (c) are similar to that in (a), but with $\kappa_0=4\kappa_{\pi}=0.0008$.}
  \label{fig5}
\end{figure*}
We now consider an asymmetric input scenario where $\kappa_ 0=4\kappa_{\pi}$.
Figure \ref{fig4}(d) shows the evolution of $\tilde{\nu}_{21-}$, $\tilde{\nu}_{12-}$, and $\tilde{\nu}_{22-}$ as a function of interaction time.
In contrast to symmetric input, the most significant difference at the 3rd-order moments of the two-mode non-Gaussian state generated by asymmetric input is that both $\tilde{\nu}_{21-}$ and $\tilde{\nu}_{12-}$ are negative within specific parameter intervals.
This indicates the presence of entanglement between the linear quadrature of mode $\hat{a}_1$ and the 2nd-order quadrature of mode $\hat{a}_2$, and vice versa.
We refer to this class of entangled states as symmetric nonlinear entangled states.
Note that the parameter interval of $\tilde{\nu}_{21-}<0$ in the case of asymmetric input is larger than that of symmetric input.
For both the 5th- and 7th-order criteria, $\tilde{\nu}_{kl-}$ is greater than 0, as shown in Fig. \ref{fig4}(e).
At the 6th-order moments, $\tilde{\nu}_{42-}$, $\tilde{\nu}_{24-}$, $\tilde{\nu}_{51-}$, and $\tilde{\nu}_{15-}$ are all negative within partial parameter intervals, whereas $\tilde{\nu}_{33-}$ remains negative across the entire parameter range, as shown in Fig. \ref{fig4}(f), suggesting the presence of five different entanglement structures.

Figures \ref{fig4}(g), \ref{fig4}(h), and \ref{fig4}(i) depict the evolution of $\tilde{\nu}_{kl-}$ as a function of interaction time for the case $\kappa_ 0=\kappa_{\pi}/4$.
In comparison to the previous two input scenarios, the parameter interval for which $\tilde{\nu}_{21-}<0$ is the largest in this configuration.
The remaining evolution law are similar to the previous asymmetric input case, and thus will not be reiterated here.
By comparing these three distinct input scenarios, we confirm that symmetric input gives rise to nonlinear entanglement, whereas asymmetric input yields symmetric nonlinear entanglement.

We next consider quantum steering between modes $\hat{a}_1$ and $\hat{a}_2$, given that steering is not only a key tool for verifying entanglement in one-sided device-independent scenarios \cite{PhysRevLett.98.140402,tian2023certification}, but also an indispensable resource for one-sided device-independent-based quantum information protocols \cite{kumar.pra.100.052329,PhysRevA.90.052325,armstrong.np.11.2015,PhysRevLett.128.200401}.
Here, we employ the Hillery-Zubairy steering criterion, which was originally proposed by Hillery and Zubairy for entanglement detection \cite{mm.prl.96.050503.2006,hillery.pra.81.062322.2010} and subsequently extended to the Einstein-Podolsky-Rosen steering criterion for linear quadratures \cite{PhysRevA.84.032115,PhysRevA.86.023626}.
For a two-mode non-Gaussian state, the Hillery-Zubairy steering criterion states: if the inequality HZ$_{kl}=|\langle\hat{A}_1\hat{A}_2\rangle|-\sqrt{\langle(\hat{A}_1^\dag\hat{A}_1+\hat{A}_1\hat{A}_1^\dag)/2\rangle\langle\hat{A}_2^\dag\hat{A}_2\rangle}>0$ is satisfied, the $l$th-order moments of mode $\hat{a}_2$ can be nonlocally steered by mode $\hat{a}_1$ through $k$th-order amplitude-and-phase quadrature measurements, where $\hat{A}_1=\hat{a}_1^k-\langle\hat{a}_1^k\rangle$ and $\hat{A}_2=\hat{a}_2^l-\langle\hat{a}_2^l\rangle$.

Figure \ref{fig5}(a) shows the evolution of HZ$_{kl}$ with $t$ under the condition of equal input power, where $\kappa_0=\kappa_{\pi}$.
In partial parameter intervals, HZ$_{42}>0$, indicating that the 4th-order quadratures of mode $\hat{a}_1$ can be steered by the 2nd-order quadrature operators $\hat{x}^2_2$ and $\hat{y}^2_2$ of measurement mode $\hat{a}_2$.
More interestingly, HZ$_{21}$ and HZ$_{51}$ are both greater than 0 in certain parameter intervals, which means the 2nd-order and 5th-order quadratures of mode $\hat{a}_1$ can be simultaneously steered via the linear quadrature operators $\hat{x}^1_2$ and $\hat{y}^1_2$ of measurement mode $\hat{a}_2$--a phenomenon not observed in the quantum state generated in Ref. \cite{zhang.pra.033701.2025}.
This unique steering property may offer distinct advantages in quantum information protocols such as high-dimensional quantum key distribution \cite{PhysRevApplied.15.034003}.
Besides, HZ$_{kl}\leq0$ ($k+l\leq7$) across the entire parameter interval, so we omit them here.
Note that for $\kappa_0=\kappa_{\pi}$, no parameter interval was found where mode $\hat{a}_1$ can steer mode $\hat{a}_2$.

Figure \ref{fig5}(b) illustrates the evolution of HZ$_{21}$ versus $t$ for the asymmetric input case with $\kappa_0=4\kappa_{\pi}$.
HZ$_{21}$ is greater than 0 only within the parameter interval $21<t<25$, indicating that the linear quadratures of mode $\hat{a}_2$ can be steered via the 2nd-order quadrature operators $\hat{x}^2_1$ and $\hat{y}^2_1$ of measurement mode $\hat{a}_1$.
In particular, the parameter interval where mode $\hat{a}_1$ can steer mode $\hat{a}_2$ is smaller than the parameter interval where the inseparability criterion $\tilde{\nu}_{21-}<0$ holds, which further indicates that quantum steering is more stringent than inseparability.
Across the entire parameter interval, all other criteria HZ$_{kl}$ remain less than 0, so we omit them here.
Figure \ref{fig5}(c) shows the evolution of HZ$_{21}$  and HZ$_{51}$ as a function of $t$, both of which are greater than 0 in some parameter intervals.
In particular, the region where they exceed 0 is significantly reduced compared to the equal-power input case, meaning that the steering properties of the two output modes at the 3rd- and 6th-order momenta can also be dynamically adjusted by tuning the input light intensity.

\begin{figure*}[htpb]
\centering
  \includegraphics[width=17.5cm]{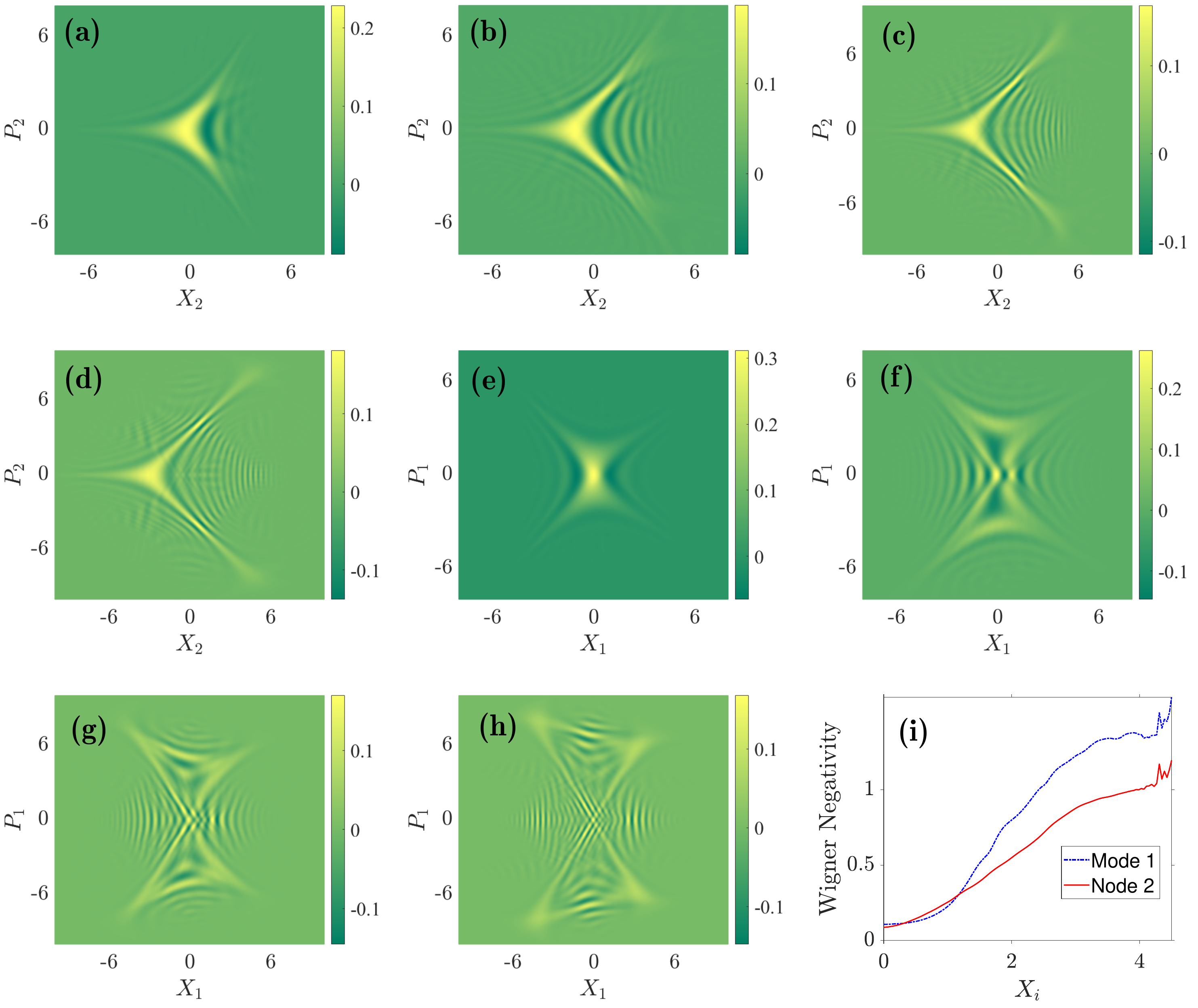}  
  \caption{The Wigner function of mode $\hat{a}_2$ is presented for measurement outcomes $x_1=1$ (a), $x_1=2$ (b), $x_1=3$ (c), and $x_1=4$ (d). (e)-(h) are similar to (a)-(d) in presentation, but correspond to the Wigner function of mode $\hat{a}_1$. The Wigner negativity of modes $\hat{a}_1$ and $\hat{a}_2$ evolves with the measurement outcome $x_i$.}
  \label{fig6}
\end{figure*}

Finally, we apply the generated two-mode state to remotely prepare Wigner-negative states via amplitude projection measurements.
In the Fock space, the amplitude projection operator can be expressed as $\hat{U}=|x\rangle\langle x|$, where
\begin{align}\label{eq9}
|x\rangle=\sum^{\infty}_{n=0}\frac{1}{\pi^{1/4}}\frac{1}{\sqrt{2^nn!}}\exp(-x^2/2)H_n(x),
\end{align}
and $H_n(x)$ denotes the $n$th-order Hermitian polynomial.

Figures \ref{fig6}(a)-\ref{fig6}(d) depict the Wigner function of mode $\hat{a}_2$ when projective measurements (with results 1, 2, 3, and 4, respectively) are performed on mode $\hat{a}_1$.
As $x_1$ increases, the profile of $W(X_2,P_2)$ evolves from its initial three-fold symmetric star shape to an asymmetric star exhibiting uniaxial symmetry.
In particular, the center of this star-shaped profile gradually shifts toward the negative $X_2$-axis with increasing $x_1$, while the interference fringes on the right side of the star become increasingly prominent.
Figures \ref{fig6}(e)-\ref{fig6}(h) illustrate $W(X_1,P_1)$ when projective measurements (with results 1, 2, 3, and 4, respectively) are conducted on mode $\hat{a}_2$.
As $x_2$ increases, $W(0_1,0_1)$ evolves from its initial maximum to the center of the interference fringes, with four vertices surrounding this center gradually emerging.
Owing to the inherent asymmetry between the output modes, the configuration of the prepared Wigner-negative states differs depending on which mode is subjected to measurement.
Furthermore, we investigate the evolution of Wigner negativity with measurement results for both scenarios, as presented in Fig. \ref{fig6}(i).
Notably, the Wigner function of the unmeasured mode remains negative regardless of the measurement result of $x_i$.
In particular, the Wigner negativity of mode $\hat{a}_1$ is markedly stronger than that of mode $\hat{a}_2$ across most of the parameter interval.

\section{Conclusion}
In summary, we found that the 3rd-order amplitude and phase quadrature operators of the fully degenerate triple-photon state generated by three-mode spontaneous parametric down-conversion exhibited alternating squeezing and anti-squeezing phenomena as the pump phase was varied.
By selecting different pump phases, four distinct fully degenerate triple-photon states were extracted, sequentially denoted as 0-phase, $\pi$/2-phase, $\pi$-phase, and 3$\pi$/2-phase squeezed states.
The nonGaussianity and nonclassicality of the system were quantified using quantum relative entropy and Wigner negativity, respectively, and both were found to increase with increasing interaction strength.
Using these squeezed states, we studied a typical scenario for generating two-mode non-Gaussian state by mixing the 0-phase and $\pi$-phase squeezed states on a beam splitter.
By employing the positive partial transpose criterion based on the higher-order covariance matrix, we characterized the entanglement of the generated state in terms of 3rd- and 6th-order moments.
We found that nonlinear entanglement could be generated under symmetric input intensities, whereas symmetric nonlinear entanglement arose under asymmetric input intensities.
Using the Hillery-Zubairy steering criterion, we characterized the quantum steering properties of the two output modes at the 3rd- and 6th-order moments.
In particular, we found that the 2nd- and 5th-order moments of mode $\hat{a}_1$ could be simultaneously steered by measuring the linear quadrature of mode $\hat{a}_2$--a phenomenon not previously reported.
This unique property holds potential for applications in high-dimensional quantum key distribution \cite{PhysRevApplied.15.034003,PhysRevA.94.052323} and quantum error correction \cite{PhysRevLett.102.120501}.
Finally, we showed that the resulting two-mode non-Gaussian state could be utilized to remotely prepare Wigner-negative states with tunable Wigner negativity.
Our results elucidate the nonclassical characteristics of fully degenerate three-photon states, thereby establishing a solid foundation for the deterministic generation of non-Gaussian entangled states using these quantum resources.
\section*{Acknowledgement}
This work was supported by the National Natural Science Foundation of China (12204293) and the Applied Basic Research Program in Shanxi Province (202203021212387).

\bibliography{my_refs_library}

\end{document}